# Analytical Modeling and TCAD simulation of n-Fz Double Sided Si Microstrip Detector Equipped with WGR irradiated by Protons for the R3B Experiment

Puspita Chatterjee[1], Nitu Saini[1] and Ajay Kumar Srivastava[1*]

[1] Department of Physics, University Institute of Sciences, Chandigarh University, Mohali, Punjab, India

E-mail: kumar.uis@cumail.in

## Abstract

Radiation hard n-Fz Double Sided Silicon microstrip Detectors are used at the Silicon Tracker for the detection of two-dimensional position and energy measurement of the incident protons in the R3B experiment at FAIR, Darmstadt, Germany.

For the development of the detectors for the R3B Silicon Tracker, the macroscopic analysis is conducted on the test structure of n-Fz Double Sided Silicon microstrip Detector, which was fabricated by BEL, Bengaluru, India, and the SRH results on the non-irradiated test structure detectors are compared with the experimental data. The SRH and CCE modeling is used to extrapolate the results up to the proton fluence of $8 \times 10^{14}$ $n_{eq}$ cm$^{-2}$ for the proton irradiated detectors. This experience is used in the designing of the detectors for the phase 1 upgrade of the experiment and the proposed Double Sided Silicon microstrip Detector equipped with Wider Guard Ring design is simulated by 2-D Silvaco ATLAS device TCAD. To evaluate the breakdown performance of the proton irradiated detectors, the inner and the outer sides (towards the cut edge of the detector) of the detector are simulated to extract the electric field distribution up to an applied bias of -1000 V. In order to observe the effect of the interstrip capacitances on the noise of the readout system of the proton irradiated (back side) detectors, SPICE simulation is performed.

The results reveal that a new radiation hard 200 μm ac coupled Double Sided Silicon microstrip Detector (with an intra guard ring) having an outer edge wider guard ring (dead space of 450 μm) structure has been proposed for the phase 1 upgrade of the R3B Silicon Tracker.

## 1. Introduction

In the NuSTAR (Nuclear Structure Astrophysics and Reaction) collaboration, the R3B (Reactions with Radioactive Relativistic Beams) experiment at FAIR (Facility for Antiproton and Ion Research) is currently under construction in Darmstadt, Germany. It was previously planned that from 2024 onwards phase 1 of the experiment will commence as part of FAIR MSV (Modularized Start Version) [1-4]. The efficiency of the Silicon Tracker (placed closest to the target region) in phase 1 of the R3B experiment, is critically dependent on the overall electrical performance attributes of proton irradiated n-Fz Double Sided Silicon micro Strip Detectors (DSSSDs) [5-6]. Among various types of silicon microstrip detectors, DSSSD stands as a valuable tool to measure the two-dimensional position of the proton track in a single detector layer, providing enhanced spatial resolution and improved tracking accuracy [7]. However, proton irradiation impacts on the detector's electrical performance by generating defect energy levels in the detector [8-9]. These trap energy levels function as charge carrier trapping centres, leading to changes in effective doping concentration ($N_{eff}$) and an increase in full depletion voltage ($V_{FD}$). Additionally, the formation of defect energy levels contributes to a rise in leakage current ($I_L$) and degradation of charge collection efficiency (CCE) under a proton irradiation environment [5]. Such performance degradations are of particular concern for the Silicon Tracker within the R3B experiment, where precise particle tracking is essential for experimental accuracy.

The impact of the defects on the performance characteristics of the n-Fz DSSSDs under proton irradiation, SRH (Shockley-Read-Hall) generation/recombination statistics in conjunction with CCE modeling is utilized [5]. These models incorporate experimentally validated proton radiation damage parameters, enabling a comprehensive evaluation of the detector's behavior under the specific proton fluence anticipated in phase 1 of the R3B experiment.

To maintain stable electrical performance in phase 1 of the R3B Experiment, the development of a radiation hard n-Fz DSSSD equipped with wider guard rings (WGR) is imperative. A critical aspect of this design involves analyzing the breakdown performance by investigating the electric field distribution profiles within the detector, and near to cut edge of the device structure by using 2-D TCAD (Technology Computer-Aided Design) simulation tool [10].

To analyse the propagation of a signal from the (back side) electrode of an n-Fz DSSSD to the readout electronics, the SPICE simulation tool is an essential component [11]. A crucial focus in an effective readout of a signal from the proton irradiated detectors is to maintain a high signal to noise ratio (SNR) to ensure accurate signal processing from the detectors [11-14]. The influence of the interstrip capacitances on the readout signal of the proton irradiated detector has been studied for the optimal performance of the detector utilized in the R3B silicon tracker.

In this paper, section 2 provides an overview of the R3B Experiment at FAIR, and section 3 describes the test structure architecture. Section 4 explains the irradiation strategy and proton radiation damage model. Section 5 delves into the current -voltage (I-V(T), capacitance-voltage (C-V ($V_{FD}$)) modeling, including SRH, CCE and SPICE modeling, which are utilized to describe the performance characteristics of the DSSSD under proton irradiation conditions. Section 6 covers the performance behaviors of the non-irradiated test structures, and section 7 describes the performance of the test structures in the proton irradiation environment. Section 8 shows the proposed device architecture of the detectors for the phase 1 upgrade of the experiment. Section 9 explains the 2D TACD simulation results on the proton irradiated detectors (inner and cut edge) for the phase 1 upgrade of the R3B experiment. Eventually, conclusions are drawn in section 10.

## 2. R3B Experiment at FAIR

The R3B experiment, a branch of NuSTAR scientific collaboration, is designed to provide kinematically complete measurement of nuclear reactions with highly energetic RIBs (Radioactive Ion Beams) up to 1.5AGeV at the phase 1 of the NUSTAR. In this experiment, FAIR is utilized as an accelerator complex, located at Darmstadt, Germany. In phase 1 upgrade of the NUSTAR experiment, the addition of double ring heavy ion synchrotron (SIS100/300) with the previous accelerator facility (GSI facility) will significantly enhance the FAIR infrastructure, enabling the production and delivery of high-energy secondary RIBs to the experimental setup [5].

In the R3B experimental setup, the high energetic RIBs interact with the fixed target, leading to emission of various particles such as protons, heavy ions, gamma rays etc. The Silicon Tracker, positioned in close proximity to the target region, serves as a crucial component for precise trajectory reconstruction of charged particles. It is composed with three layers: one inner layer and two identical outer layers. The inner layer consists of B and D sensors, integrated with 24 ASICs (Application Specific Integrated Circuits) per layer while the outer layer is constituted with A, B and C sensor, incorporating 32 ASICs per layer. The overall performance of the Silicon Tracker highly relies on the operational characteristics of the DSSSDs. The inner layer of the Silicon Tracker is composed with six 100 μm n-Fz DSSSDs, whereas each outer layer comprises twelve 300 μm n-Fz DSSSDs.

The performance of n-Fz DSSSDs with a thin 100 μm n-Fz DSSSD in the inner layer under Phase 1 irradiation conditions is acceptable as only reasonable values of $V_{FD}$, $I_L$, and CCE are observed. However, the 300 μm DSSSDs in the outer layer are severely degraded after irradiations, indicated by excessive $V_{FD}$ increase and substantial CCE decrease. This severe radiation damage compromises their operational efficiency, necessitating the replacement of the outer layer 300 μm n-Fz DSSSDs with a more radiation-hard detector design to ensure reliable performance under phase 1 high proton fluence conditions [5].

### 3. Test structure architecture

The n-Fz DSSSD test structure was fabricated at BEL (Bharat Electronics Limited), Bengaluru, India by Tata Institute of Fundamental Research (TIFR), Mumbai [15] and the device is used here as a sample test structure for the designing and development of the proposed DSSSD detector (with an intra guard ring structure) equipped with WGR for the phase 1 upgrade of the R3B experiment. Figure 1(a) shows the 2-D visualization of the schematic cross-section of a 300 μm thick test structure n-Fz DSSSD for the SRH and CCE modeling study. The resistivity of the n-Fz <100> Si material is 10 kΩ-cm, which is equivalent to the doping concentration of $4.62 \times 10^{11}$ cm$^{-3}$.

The frontside of the detector is implanted with n$^+$ strips, each having a width of 12 μm, pitch of 50 μm and length of 25600 μm. These 1024 n$^+$ strips are critical for the collection of electrons generated by incident charged particles. To prevent electrical crosstalk and to ensure accurate position measurement, the n$^+$ strips on the frontside of the detector is electrically isolated using a p-channel stop with a width of 7 μm. On the backside of the detector, 512 p$^+$ strips are implemented, each having a width of 50 μm, pitch of 75 μm and length of 76800 μm. These p$^+$ strips, orthogonal to n$^+$ strips, serve to collect holes, generated within the silicon bulk. The configuration and implantation of these strips are designed to complement the frontside n$^+$ strips, thereby enabling the detector to capture two-dimensional position information of particle tracks. To operate the n-Fz DSSSD in full depletion condition, DC reverse bias voltage is applied: the backside is biased with a negative potential while the frontside is kept as grounded.

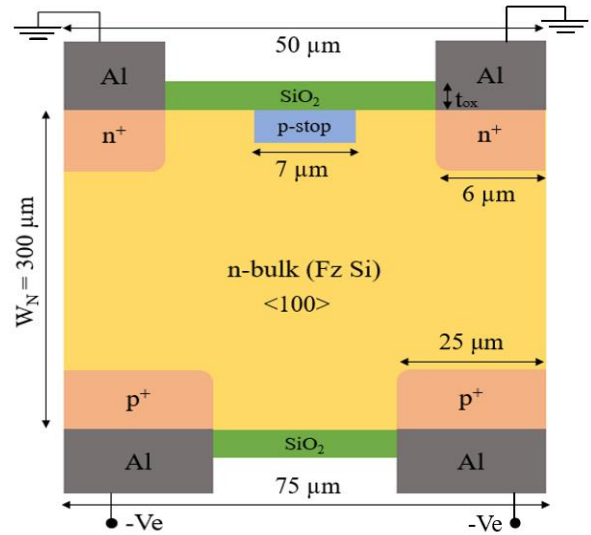

**Figure 1.** Schematic cross-section of a DC coupled n-Fz DSSSD test structure.

### 4. Irradiation strategy and Proton irradiation damage model

When the n-Fz DSSSD is exposed to proton irradiation, defect energy levels introduced in the silicon bulk, lead to degrade the electrical performance of the detector. DLTS (Deep Level Transient Spectroscopy) and TSC (Thermally Stimulated Current) techniques are commonly employed by the research community to characterize this radiation induced defect energy levels. These techniques provide crucial insight into the behavior of the defect energy levels. The radiation damage models incorporate critical parameters: such as concentration of the defect energy levels, energy levels of the defect and the electron and hole capture cross-sections. Table 1 shows the proton radiation damage model [16], that is used for the SRH, CCE and TCAD device simulation to study the impact of radiation damage effects on the n-Fz DSSSD.

**Table 1.** Proton irradiation damage model for radiation damage analysis [16].

| Trap levels | Energy level of traps (eV) | Capture cross-section of electrons (cm²) | Capture cross-section of holes (cm²) | Concentration of trap levels (cm⁻³), φ is an equivalent fluence (1 MeV neutron) |
|---|---|---|---|---|
| Deep acceptor | $E_C$-0.525 | $1\times10^{-14}$ | $1\times10^{-14}$ | $1.189\times\phi$ cm⁻³ $+0.65\times10^{14}$ |
| Deep donor | $E_V$+0.48 | $1\times10^{-14}$ | $1\times10^{-14}$ | $5.589\times\phi$ cm⁻³ $-3.96\times10^{14}$ |

## 5. Current-voltage, Capacitance-Voltage ($V_{FD}$), SRH and CCE modeling

According to the Shockley-Read-Hall (SRH) modeling,

$$e_{n,p} = C_{n,p} N_{C,V} \exp\left(\pm \frac{E_t - E_{C,V}}{K_B T}\right) \quad (1)$$

Here, the emission rates for electrons $e_n$ and holes $e_p$ depend on parameters such as the trap energy level $E_t$ and electron capture cross-sections $\sigma_n$, hole capture cross-sections $\sigma_p$, while the densities of states $N_c/N_v$ also play crucial roles and the $C_{n,p}$ is the capture cross-section of electron/hole.

$N_{eff}$ can be expressed as [5]:

$$N_{eff} = N_D + \left(\sum_{donor} p_t - \sum_{acceptor} n_t\right) \quad (2)$$

Here, $N_D$ represents the initial doping concentration, and $n_t/p_t$ denotes the population of trap levels by electrons and $p_t$ denotes the population of trap levels by holes, respectively. These defect levels influence the detector's electrical properties, contributing to variation in leakage current, which is calculated as [5]:

$$I_L = qAW_N\left(\sum_{donor} p_t e_p + \sum_{acceptor} n_t e_n\right) \quad (3)$$

Here, q is the charge of electron, A is the detector's active area, and $W_N$ is the thickness of the detector in full depletion condition.

This relationship links the detector's effective doping concentration to the voltage required to fully deplete the active region, a critical factor in understanding its performance post-irradiation.

To evaluate the detector's charge collection efficiency (CCE) under proton irradiation, the following CCE modeling is employed. The CCE, defined as the ratio of charge collected by the segmented electrode ($Q_{in}$) to the total charge produced ($Q_0$) by the detector, is given by [5]:

$$CCE = \frac{Q_{in}}{Q_o} = \frac{W_N}{W} \frac{\tau_{eff}}{t_{dr}}\left(1 - e^{-\frac{t_{dr}}{\tau_{eff}}}\right) \quad (4)$$

Here, $W_N$ is the depleted width at a particular bias, W is the device depth (at V=$V_{FD}$, W=$W_N$), and $t_{dr}$ represents the drift time of carriers. The effective carrier lifetime $\tau_{eff}$ is inversely proportional to the defect density ($N_t$) [5].

## 6. Performance characteristics of test structure in non-irradiation environment

To understand the performance of the test structure, firstly, the I-V and C-V measurement is performed on the non-irradiated structures. Figure 2 demonstrates the I-V characteristic of the p⁺ strip (back side) of the non-irradiated test structure at room temperature (T=300 K). In order to compare the results of the SRH modeling on $I_L$, and $V_{FD}$, the experimental data is compared with the theoretical result with an uncertainty of ±10%, which is due to the temperature differences in the probing area of the detector during measurement [17]. The initial increment of current in the I-V curve per strip is due to the diffusion of charge carriers within the reverse-biased detector. At full depletion conditions, the reverse saturation current corresponds to the movement of minority charge carriers generated by thermal energy. It can be evident from figure 2 that the theoretical and experimental curve shows close alignment up to a point of full depletion. However, after the full depletion condition, the modeling, which considers W=$W_N$ at $V_{FD}$, predicts a saturation region in the I-V curve per strip. Conversely, the experimental curve shows an electric field enhancement effect beyond full depletion and also it may be due to the surface current from the detector's depleted surface at the Si-SiO₂ interface of the (back side) detector, resulting in a divergence from the modeling predictions. Notably, despite these deviations, the experimental results and the theoretical predictions remain within the ±10% experimental uncertainty range, underscoring the robustness of the experimental data.

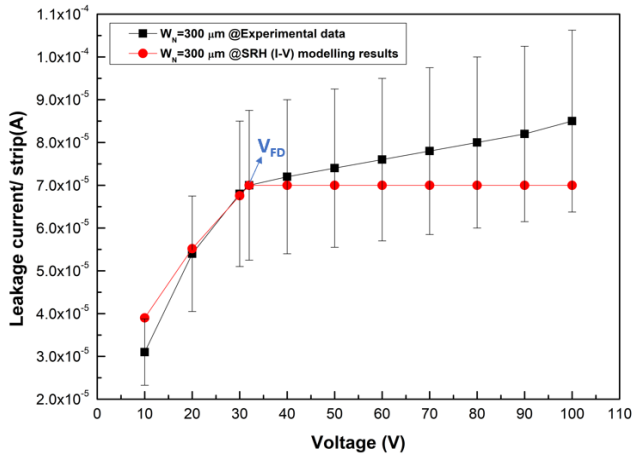

**Figure 2.** Current-voltage characteristics of the test structure in the non-irradiation environment at 300 K.

A very good agreement is observed at I-V at 300 K/C-V ($V_{FD}$=33V) between experimental data [15] and SRH modeling results.

## 7. Performance characteristics of test-structure in proton irradiation environment

While the detector is exposed to proton irradiation, defect energy levels are introduced (see table 1) into the detector and act as trapping centres for charge carriers. These defect energy levels modify the overall space charge distribution of the bulk, leading to a change in the $N_{eff}$, which subsequently raises the $V_{FD}$ and $I_L$ while also decreasing CCE. To evaluate the test structure's performance under proton irradiation, a proton radiation damage model is utilized. The microscopic radiation damage parameters from experimentally validated proton irradiation damage model were subsequently integrated into the SRH and CCE modeling framework (described expressed in equations (2), (3) and (4)) to predict the variation of $V_{FD}$, $I_L$ and CCE with increasing proton fluence.

The radiation damage model from table 1 utilized in these studies introduces one acceptor and one donor trap energy levels in between the middle of the bandgap and conduction/valence band of the detector. Figure 3(a) illustrates the increase in $V_{FD}$ for the DSSSD test structure as a function of proton irradiation fluence from $1\times10^{14}$ $n_{eq}$ cm$^{-2}$ to $5\times10^{14}$ $n_{eq}$ cm$^{-2}$. As depicted in the figure, at a fluence of $1\times10^{14}$ $n_{eq}$ cm$^{-2}$, the $V_{FD}$ is observed at 212 V while at a fluence of $5\times10^{14}$ $n_{eq}$ cm$^{-2}$ the $V_{FD}$ is noticed to reach approximately 807 V with an experimental uncertainty of ±10% [17]. This is because of the incorporation of two defect energy levels in the detector, changes in overall space charge within the detector leads to an increase of $N_{eff}$ and subsequently an increase in the $V_{FD}$ of the n-Fz DSSSD.

Since the defect energy levels are positioned within the middle of the bandgap and conduction/ valence band, it can influence significantly to the overall space charge and as well as it can also contribute to the $I_L$ within the detector. Figure 3(b) illustrates the increase in $I_L$ with an increasing proton irradiation fluences from $1\times10^{14}$ $n_{eq}$ cm$^{-2}$ to $5\times10^{14}$ $n_{eq}$ cm$^{-2}$ with an experimental uncertainty of ±10%. These defect energy levels can also function as trapping centres, capturing charge carriers and thereby reducing the number of carriers available for collection at the junction. The defect energy levels trapped the charge carriers, and there was a notable loss of drifting charge carriers collected by the electrode, leading to a decrease in the CCE of the detector with an experimental uncertainty of ±5% [5]. Figure 3(c) illustrates the degradation of CCE as a function of proton irradiation fluence. CCE is observed to be around 90% at a fluence of $1\times10^{14}$ $n_{eq}$ cm$^{-2}$, while CCE decreases to approximately 60% at a higher fluence of $5\times10^{14}$ $n_{eq}$ cm$^{-2}$.

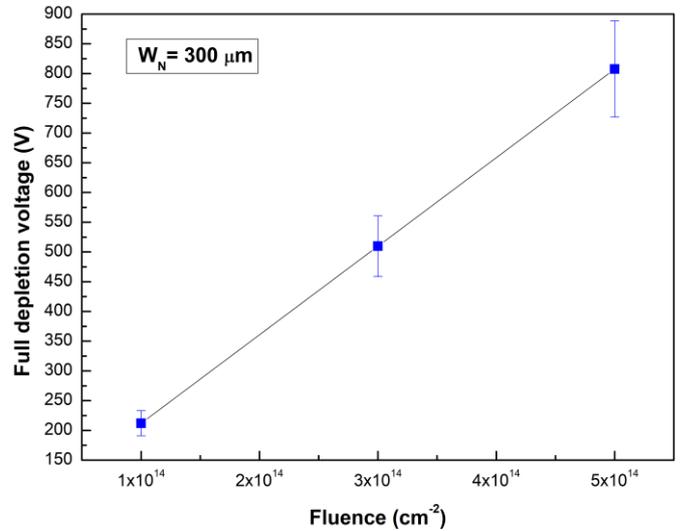

(a)

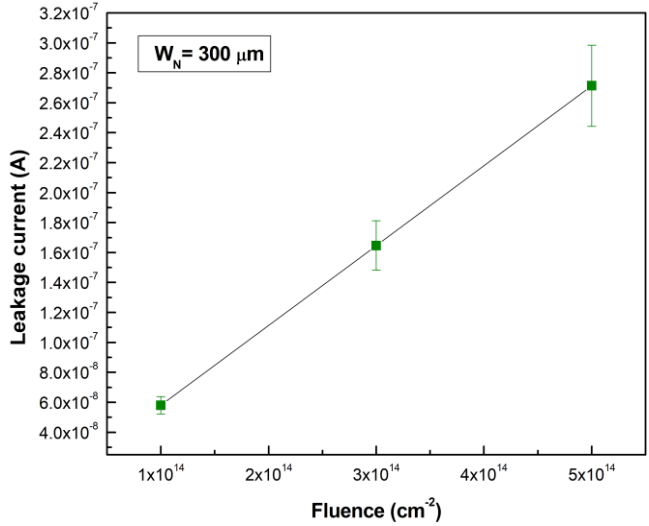

(b)

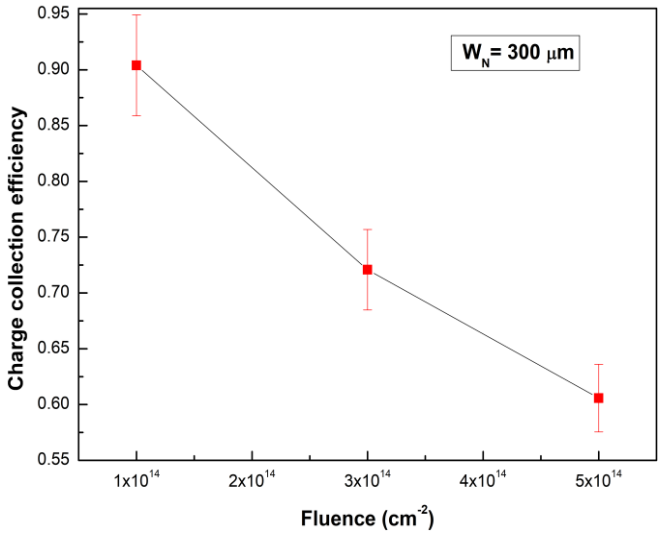

(c)

**Figure 3.** SRH and CCE modeling of a proton irradiated test structure **(a)** Full depletion voltage **(b)** Leakage current **(c)** Charge collection efficiency as a function of proton fluences.

## 8. Proposed device architecture for the Phase 1 of the R3B experiment

The 200×50 µm$^2$ of ac coupled n-Fz DSSSD design (sensor A, B and C for the outer layer of the silicon Tracker) for the phase 1 upgrade of the R3B Experiment is depicted in figure 4(a) [5]. The corresponding device and process parameters are mentioned in table 2. The n$^+$n$^-$p$^+$ device structure consists of n$^+$ strips on the front side with a gaussian peak doping profile/ concentration of $1\times10^{16}$ cm$^{-3}$ at the surface of the detector, n-Fz Si bulk having uniformly doped at a concentration of $5\times10^{12}$ cm$^{-3}$ and p$^+$ strips on the backside exhibiting gaussian peak doping profile/ concentration of $1\times10^{18}$ cm$^{-3}$ at the Si-SiO$_2$ interface of the detector. The front side of the detector is segmented by 38 µm n$^+$ strips with a strip pitch of 50 µm, while the backside features 38 µm p$^+$ strips with the same strip pitch of 50 µm, while the frontside (n$^+$) strips and the backside (p$^+$) strips are arranged orthogonally with each other. Electrical isolation between the n$^+$ strips on the front side is achieved using a 1.4 µm deep p-spray technique with doping concentration of $3\times10^{15}$ cm$^{-3}$. The junction depth of the p$^+$ and n$^+$ side of the detector is precisely 1 µm, and the passivation layers consist of a 300 nm SiO$_2$ layer and a 50 nm Si$_3$N$_4$ layer.

On the backside of the detector, a double metal structure is incorporated, while two metals are electrically connected through a via connection to take the readout of the signal in a single direction. Additionally, an p$^+$ intra-guard ring structure with a depth of 1 µm, width of 5 µm and doping concentration of $1\times10^{18}$ cm$^{-3}$ is also implemented between two p$^+$ strips on the backside of the detector to overcompensate the electron accumulation layer (EAL).

In the device the frontside (n$^+$ strips) and backside (p$^+$) strips are biased through the bias resistor (not shown in figure 4(a), it is usually visualized in the three-dimensional view of a detector)). In the design, Dirichlet boundary conditions are applied on the contact surfaces and the Neumann boundary conditions are applied on the non-contact surfaces. In the present work, the detector is biased through the DC pad or ohmic contact (shown in red colour in figure 4(a)).

Around the periphery of the detector, guard rings are strategically implemented to enhance the breakdown voltage performance of the detector and to reduce the cut edge effect. As depicted in figure 4(b), the proposed double sided outer edge WGR design for the n-Fz DSSSD features a 120 µm expanded guard ring layout [5]. The front side of the detector features an n-type guard ring while the backside is composed of a p-type guard ring. The doping of the front side (p$^+$) and back side (n$^+$) of the WGR is same as the doping concentration of p$^+$ and n$^+$. According to the industry trend, the total inactive space is considered as 450 µm having one wider guard ring of 120 µm. In this configuration, the last p$^+$ strip of the detector is biased with negative potential, while the guard ring is kept floating and the end n$^+$ layer is biased with zero voltage.

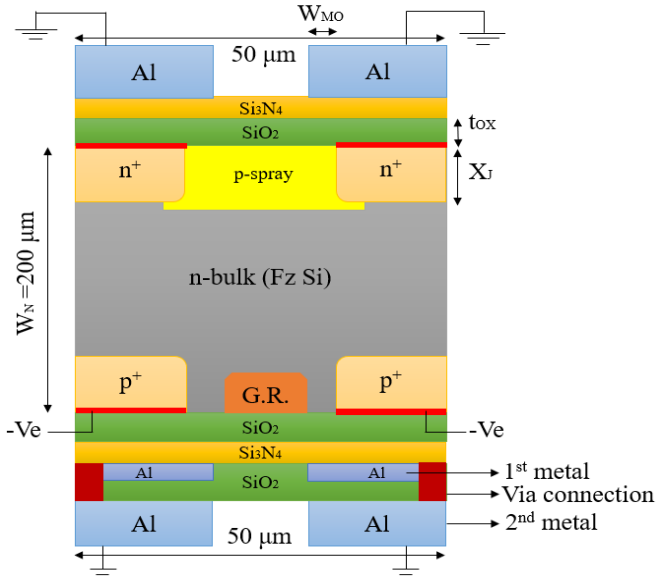

| Table 2. Device and process parameters of proposed ac coupled 200 μm n-Fz DSSSD. | |
|---|---|
| **Device and process parameters** | **Values** |
| Bulk doping type | n-type ($5\times10^{12}$ cm$^{-3}$) |
| Device thickness | 200 (Sensor A, B, C) /100 (Sensor B, D) μm |
| Strip pitch | 50 μm |
| Strip width | 38 μm |
| Thickness of SiO$_2$ ($t_{ox}$) | 300 nm |
| Thickness of Si$_3$N$_4$ | 50 nm |
| Isolation implant | p-spray (depth=1.4 μm, concentration=$3\times10^{15}$ cm$^{-3}$) |
| Junction depth ($X_J$) | 1 μm |
| Width of metal over hang ($W_{MO}$) | 5 μm |
| Intra guard ring structure | Depth= 1 μm, width= 5 μm, concentration=$1\times10^{18}$ cm$^{-3}$ |
| p$^+$ strips doping | $1\times10^{18}$ cm$^{-3}$ |
| n$^+$ strips doping | $1\times10^{16}$ cm$^{-3}$ |

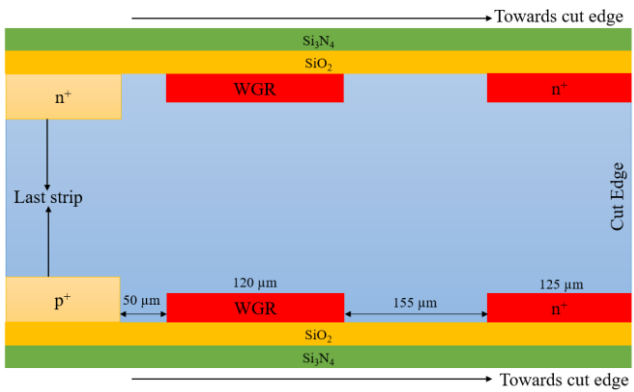

**Figure 4(a).** Schematic cross-section of proposed ac-coupled n-Fz Double Sided Silicon Strip Detector for the phase 1 of the R3B Silicon Tracker. **(b).** Schematic cross-section of proposed outer edge wider guard ring design layout with expanding width (Biased resistors are not shown in the 2D schematic).

## 9. Performance characteristics of n-Fz DSSSD (inner & outer part: equipped with WGR structure) using TCAD simulation

The 2D TCAD (Silvaco) device simulation is performed in this section to investigate the breakdown performance of the proposed n-Fz DSSSD design (shown in figure 4(a) and (b)), irradiated with phase 1 proton irradiation fluences. In the TCAD Silvaco device simulator interface the ac-coupled n-Fz DSSSD design has been constructed by using the following steps: defining the mesh structure, defining the region, defining electrodes, defining the doping profile for every particular region and so on. The device simulator involves solving the Poisson equation within the silicon bulk of the detector, as well as the continuity equations on the silicon lattice by incorporating the mobility model, Shockley–Read–Hall (SRH) generation/ recombination model, impact ionization model to accurately represent the physical phenomena within the device.

*9.1. Breakdown performance of the inner part of the n-Fz DSSSD*

To analyze the breakdown performance of the n-Fz DSSSD (as shown in figure 4(a)) irradiated by proton fluences $2\times10^{14}$, $5\times10^{14}$, $8\times10^{14}$ $n_{eq}$ cm$^{-2}$, 2-D simulation has been conducted using TCAD (Silvaco) ATLAS device simulation tool. In the case of irradiated detectors, the fixed oxide charges ($N_{ox}$) of $1.5\times10^{12}$ cm$^{-2}$ are taken at the Si-SiO$_2$ interface of the n-Fz DSSSD. The backside of the device is reverse biased up to -850V considering a 5 µm metal overhang implemented on both sides of the device.

Figures 5(a) and 5(b) show the surface electric field distribution in the detector, which is just below 300 nm from the Si-SiO$_2$ interface for the front side (n$^+$) and the back side (p$^+$) of the n-Fz DSSSD. The electric field is decreasing with an increasing fluence at the center of the detector, and also it is showing a dip at the middle of the detector i.e. x=25 µm, which is due to the increasing concentration of electron in the EAL (Electron Accumulation Layer) in the presence of p-spray layer (see figure 5(c)) with an increasing fluences. It has been observed that the surface electric field reaches a peak value of 13000 V/cm at the fluence of $2\times10^{14}$ $n_{eq}$ cm$^{-2}$ and with the increase of proton equivalent fluence the electric field progressively decreases. As the electric field at the frontside of the detector is much lower than the value of the critical electric field ($3\times10^5$ V/cm) up to the applied bias of -850 V, the detector does not show any breakdown up to -850 V.

The electric field strength at the backside of the detector reaches 6000 V/cm at the fluence of $8\times10^{14}$ $n_{eq}$ cm$^{-2}$ at the curvature of p$^+$ junction, and it decreases with decreasing fluences, the reduction in the electric field is also observed. The observed dip in the backside electric field profile of the detector is due to the incorporation of an electron accumulation layer between two consecutive p$^+$ strips at the backside of the detector. Due to the incorporation of intra guard ring (p$^+$), full compensation of the concentration of the electron in the EAL has been observed with an increasing fluence (see figure 5(d)).

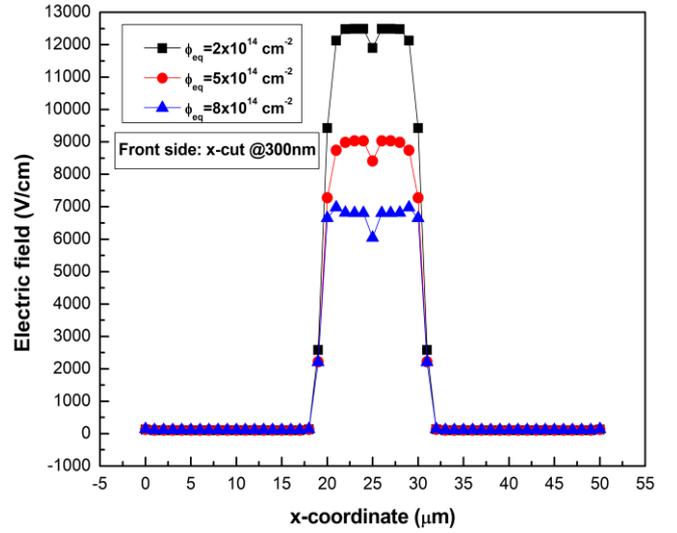

(a)

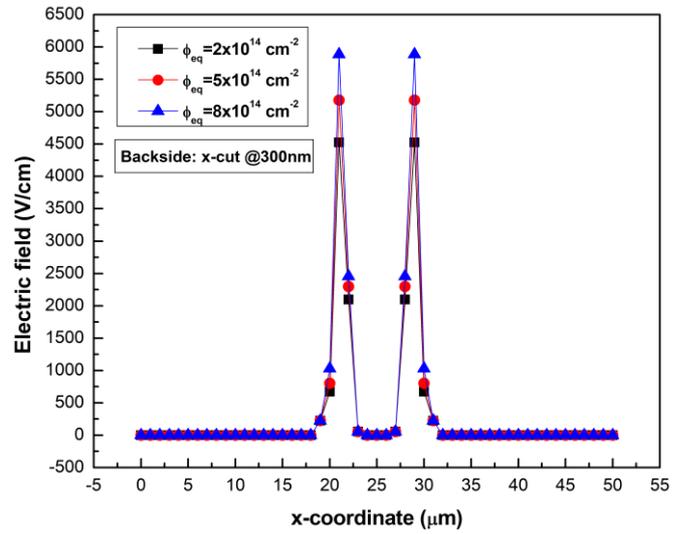

(b)

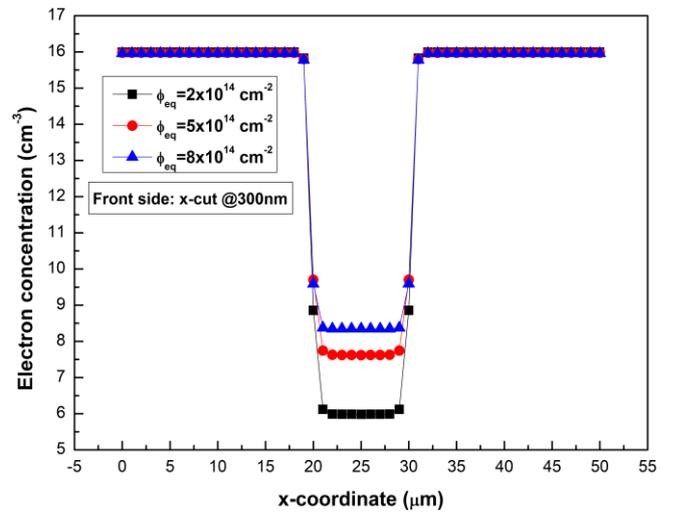

(c)

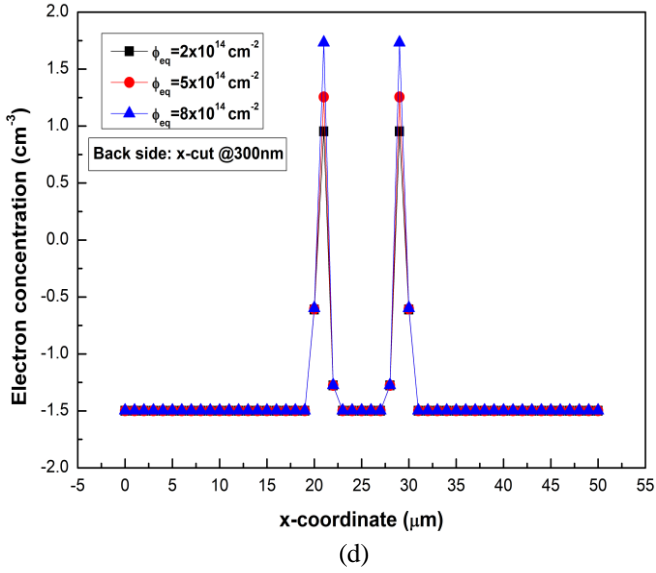

(d)

**Figure 5(a).** Electric field profile of the front side of the n-Fz DSSSD (50×200 μm$^2$), **(b).** Electric field profile of the back side of the n-Fz DSSSD, **(c).** Electron concentration profile of the front side of the n-Fz DSSSD, **(d).** Electron concentration profile of the back side of the n-Fz DSSSD.

## 9.2. Breakdown performance of outer part of the n-Fz DSSSD

The TCAD simulation is performed to extract the electric field distribution just 300 nm below the Si-SiO$_2$ interface in case of the cut edge configuration of the proton irradiated n-Fz DSSSD (see figure 4(b)), by incorporating a single WGR of 120 μm. The device featured 5 μm of metal overhang and applied reverse bias up to -1000 V to plot the surface electric field distribution at a bias voltage of -1000 V. In the simulation, the WGR is considered as a floating electrode, n$^+$ implanted anodes were grounded at zero bias and the p$^+$ implanted strip is biased up to -1000 V. Figure 6(a) and 6(b) show the surface electric field distribution (just 300 nm below the Si-SiO$_2$ interface) on both the front side and the backside of the outer edge of the n-Fz DSSSD for all fluences. It is noted that the maximum electric field is observed to be 3250 V/cm on the front side and 4300 V/cm on the backside in the irradiated detector (at a fluence of 8×10$^{14}$ n$_{eq}$ cm$^{-2}$), which is less than the critical electric field value of 3×10$^5$ V/cm. This indicates that the device incorporating a outer edge WGR in the optimized design can be operated up to -1000V without experiencing any avalanche breakdown. This configuration, therefore, provides enhanced operational stability under high-voltage conditions, making it suitable for high-radiation environments.

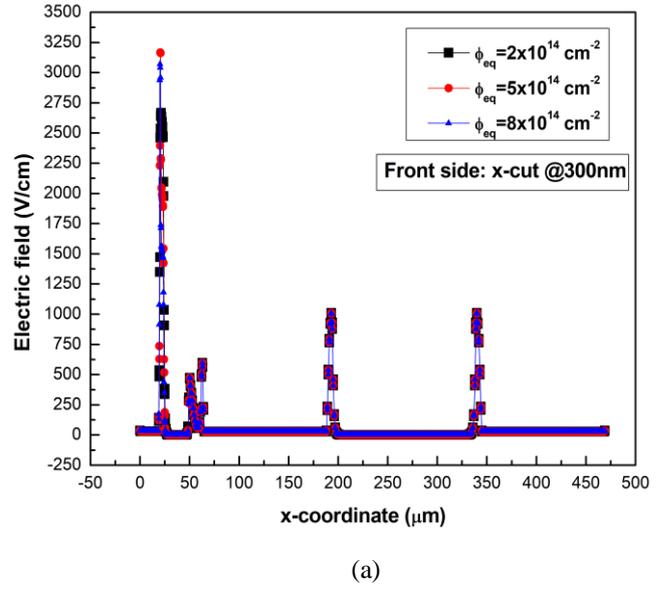

(a)

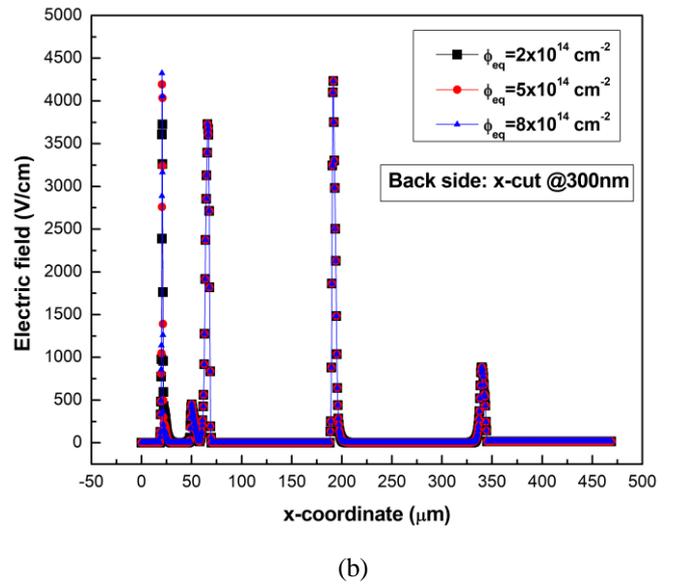

(b)

**Figure 6(a).** Surface electric field profile of 300 nm below the interface at the front side (outer edge) of the n-Fz DSSSD by introducing 120 μm WGR at the cutedge of the detector. **(b).** Electric field profile of 300 nm below the Si-SiO$_2$ interface at the back side (outer edge) of the n-Fz DSSSD by introducing 120 μm WGR at the cutedge of the detector.

## 9.3. Investigation of 1$^{st}$ stage signal propagation in proton irradiated n-Fz DSSSD using SPICE

In this section, the initial stage of signal amplification from 23 MeV proton irradiated n-Fz DSSSD is analyzed for different proton fluences and also for different detector (sum of back plane and interstrip) capacitances, utilizing TINA TI SPICE simulation software by Texas Instrument.

In figure 7, a step pulse is used at the input of the CSP (Charge Sensitive Preamplifier) which is equivalent to the charges collected per strip in the detector irradiated by different proton fluences. In Figure 7, the circuit consists of a CSP along with a shaping amplifier CR-RC-RC network is used for the detection of the signal to noise ratio as expected and to ensure the proper baseline restoration of the signal output. A pole-zero cancellation network, parallel to the $C_{diff}$ in the shaping network is incorporated, to address the undershoot and pile-up effects for maintaining a stable baseline [18].

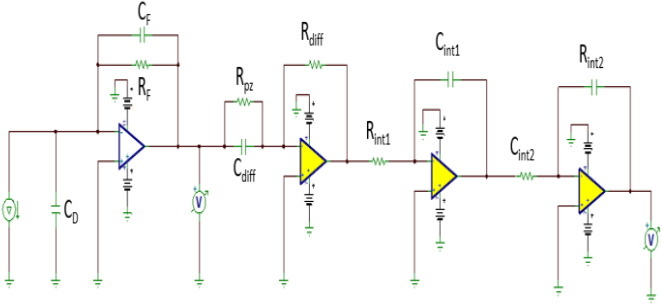

**Figure 7.** 1st stage signal propagation network from the proton irradiated n-Fz DSSSD.

Firstly, to deduce the current pulse response of the detector irradiated to three different proton fluence levels ($2\times10^{14}$, $5\times10^{14}$, $8\times10^{14}$ $n_{eq}$ $cm^{-2}$), it is very crucial to optimize the number of charges produced per μm of the detector for each proton fluence. Figure 8 shows the number of charge carriers produced per μm of the detector with the increase of proton fluences: $2\times10^{14}$ $n_{eq}$ $cm^{-2}$, $5\times10^{14}$ $n_{eq}$ $cm^{-2}$, $8\times10^{14}$ $n_{eq}$ $cm^{-2}$ and this is an approximated as per the SRH modeling of the $I_L$ for the different fluences in the irradiated detectors at the temperature of 253 K. That leads to the generation of electron hole pairs, contributing to higher charge carrier density per micron of the detector.

The number of charge carriers collected ($Q_{in}$) by the strips of the detector can be expressed as [5]:

$$CCE = \frac{Q_{in}}{Q_0} \quad (5)$$

Here, $Q_0$ is the amount of charge produced by the n-Fz DSSSD after the interaction of the radiation with the detector.

The conversion of the current pulse from the amount of charge carriers collected by the strips of the detector is expressed as:

$$Q_{in} = \int_0^t I\,dt \quad (6)$$

Here, I is the current pulse generated by the detector and t is the duration of the current pulse.

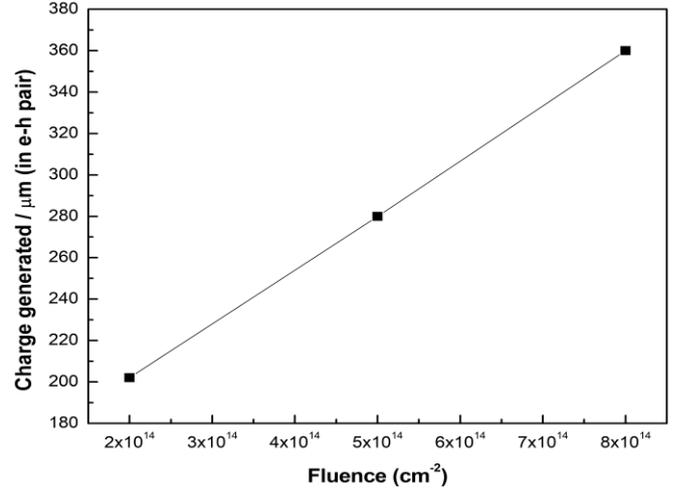

**Figure 8.** e-h pair generation per micron of the 23 MeV proton irradiated n-Fz DSSSD for the different proton fluences.

After optimizing the $Q_0$ for three different proton fluences, corresponding current pulses were determined from the eq (6). This step current pulse is processed as the input to the CSP circuit in the subsequent step having a pulse duration of 2.05 ns and amplitude of 2.60, 6.50, 10.4 μA for three proton fluences $2\times10^{14}$, $5\times10^{14}$, $8\times10^{14}$ $n_{eq}$ $cm^{-2}$ respectively.

In the next step, a CSP circuit is designed to convert input charge signals to voltage pulses by ensuring that the time constant (τ) of the CSP is much greater than the pulse duration (t) for effective charge collection. The input and output voltage formulation of the CSP is given by [12]:

$$V_{in} = \frac{Q_{in}}{C_d} \quad (7)$$

$$V_{out} = -\frac{Q_{in}}{C_F}e^{-\frac{t}{\tau}} = -\frac{\int I\,dt}{C_F}e^{-\frac{t}{\tau}} \quad (8)$$

Here, $V_{in}$ is the input voltage of the CSP, $C_d$ (= $C_J$ + $C_{int}$) is the total detector capacitance, $V_{out}$ is the output voltage of the CSP, $C_F$ is the feedback capacitance of the CSP and the $R_F$ is the feedback resistance of the CSP and $\tau=R_FC_F$ is the time constant of the CSP.

In the SPICE simulation framework, a CSP LF-444A quad low power operational amplifier (OPAMP) is considered due to its advantages of large reverse breakdown voltage and low noise voltage and low input noise current [11]. Figure 9 shows a comparison between the simulated (by SPICE) and modeled (according to eq (8)) output voltage pulses for three different proton fluences from the CSP. In the initial observation shown in figure 9(a) $C_F$ and $R_F$ are considered as 2 pF and 10 MΩ respectively, resulting in a time constant (τ) of 20 μs, which is significantly higher than the pulse duration (t) of 2.05 ns. However, in this case, the simulated results and modeled data do not closely align with each other. In the next observation (figure 9(b)), the $C_F$ and $R_F$ values are adjusted to 1 pF and 0.1 MΩ respectively, reducing the difference between the τ of 100 ns and t. This adjustment improved the agreement between the simulated results and modeled data than the previous case. In figure 9(c) the $C_F$ and $R_F$ values were optimized to 1.1 pF and 10 kΩ, obtaining a close alignment between simulated results and modeled data with an uncertainty of ±10%. These optimized parameters are taken into account for further processing of the signal pulse through the CR-RC$^2$ shaper.

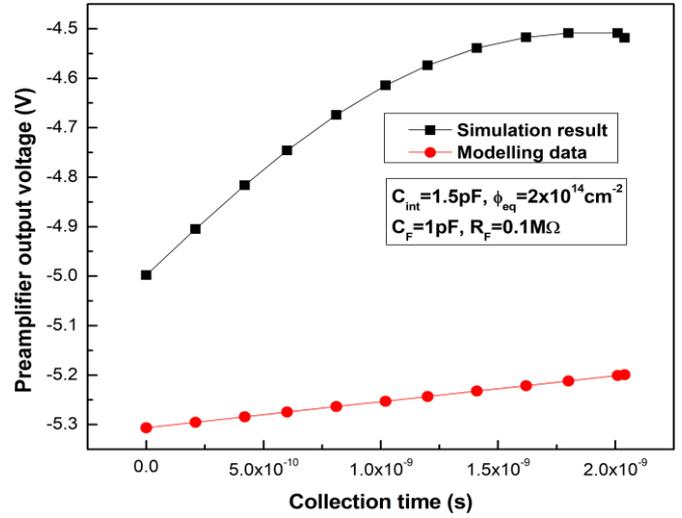

(b)

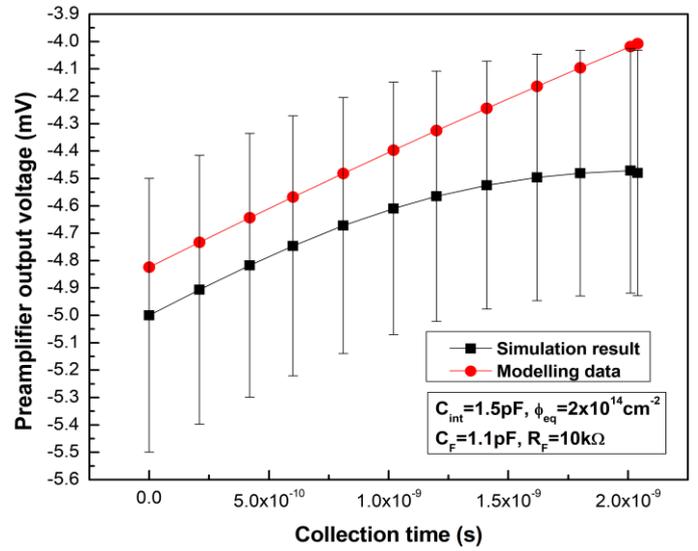

(c)

**Figure 9.** Comparison of the charge sensitive preamplifier's simulated and the modelled output pulses with **(a)** time constant >>> pulse duration, **(b)** time constant >> pulse duration, **(c)** time constant > pulse duration.

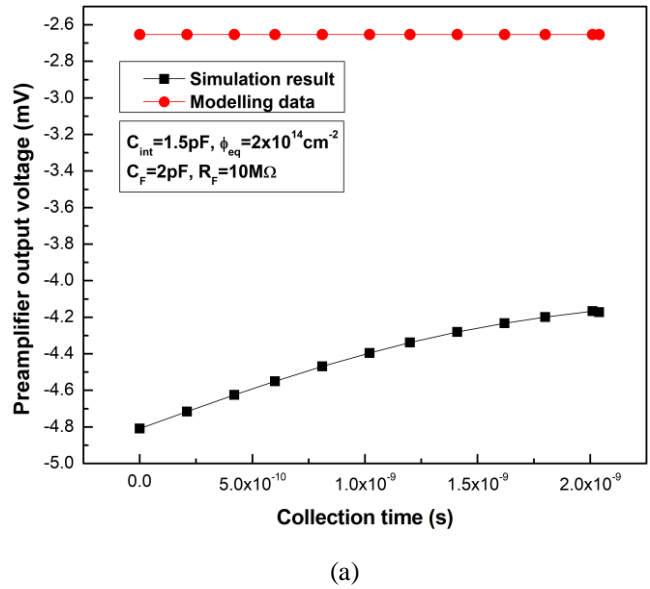

(a)

After that, a CR-RC-RC network is developed to improve signal to noise ratio and ensure proper baseline restoration. For designing CR-RC$^2$ network OPA-657 is used, for providing low distortion voltage feedback [11, 18]. Incorporation of a pole-zero cancellation network (parallel to the $C_{diff}$) has been done, to address undershoot and pile-up effects for maintaining a stable baseline. The output voltage of a simple CR-RC network is denoted by [12]:

$$V_{out} = -\frac{C_{diff}}{C_{int}} V_{in} \frac{\tau_{diff}}{\tau_{diff}-\tau_{int}} \left[ e^{-\frac{t}{\tau_{diff}}} - e^{-\frac{t}{\tau_{int}}} \right] \quad (9)$$

Here, $C_{diff}$ and $R_{diff}$ are the capacitance and resistance of the first CR differentiator network, $C_{int}$ and $R_{int}$ is the capacitance and resistance of the first RC integrator network.

Figure 10(a), (b) and (c) illustrate the influence of different interstrip capacitance values on the signal shaping for distinct three proton fluences: $2\times10^{14}$ cm$^{-2}$, $5\times10^{14}$ cm$^{-2}$, $8\times10^{14}$ cm$^{-2}$. The result shows stable signal shaping without showing any signal pile-up or baseline distortion for the different proton fluences. It can be seen from the figure that the shaper output pulse increases with the increase of proton fluences but it shows no significant differences for varying interstrip capacitance. This is due to the consideration of significantly long shaping time which can effectively reduce the high frequency noise, arising from interstrip capacitance. The shaping amplifier effectively mitigates the influence of interstrip capacitance on signal shaping by enhancing the signal to noise ratio, thereby ensuring minimal distortion in the signal.

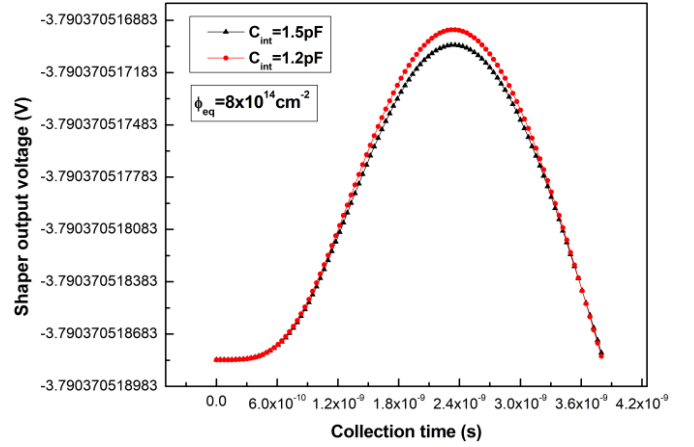

(c)

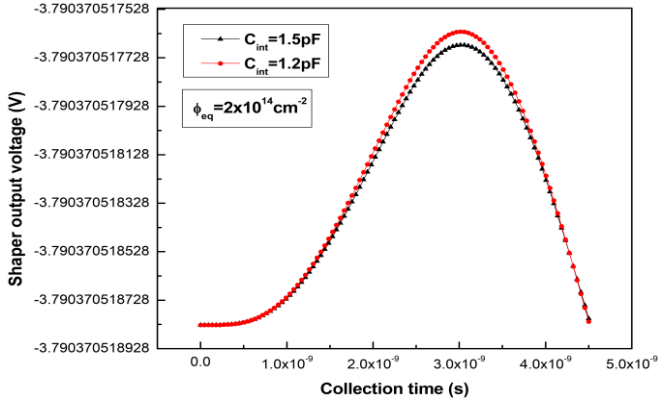

(a)

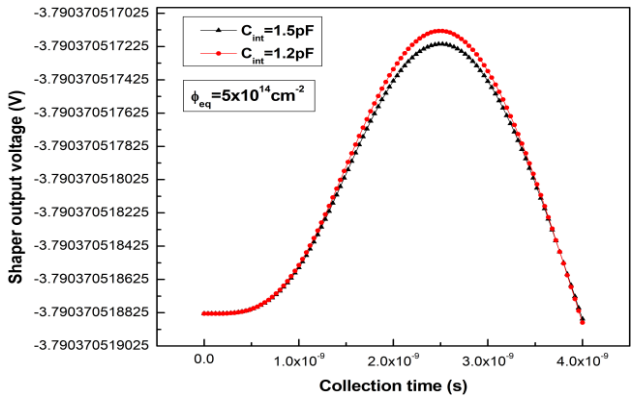

(b)

**Figure 10(a).** Comparison of shaper output pulse for two distinct different interstrip capacitance connected to the proton irradiated detector exposed with fluence of $2\times10^{14}$ cm$^{-2}$, **(b)** $5\times10^{14}$ cm$^{-2}$ and **(c)** $8\times10^{14}$ cm$^{-2}$.

## 10. Conclusion

In this contribution, a proton irradiation damage model is used which gives a significantly better and consistent description of a large set of measurements I–V(T), C–V ($V_{FD}$), and CCE of test structures and TCAD device simulation of the proposed design layout of n-Fz DSSSD detector with a p$^+$ intra guard ring and a WGR at cut edge of the detector irradiated with protons up to a fluence of $8\times10^{14}$ n$_{eq}$/cm$^2$.

The macroscopic results are extrapolated using SRH, and CCE modeling on the irradiated test structure, and the non-irradiated detector's macroscopic results are compared with the experimental data. A good agreement between measurements and simulation results on the test structure has been obtained on the $I_L$, and $V_{FD}$ at a temperature of 300 K with an uncertainty up to ±10%.

This experience is used in the designing and simulation of the detectors using Silvaco ATLAS TCAD device simulation of the detector, and it is shown that with the cut edge of < 450 μm, no avalanche breakdown is observed up to -1000 V applied bias. The signal processing of the detector (p$^+$ back side) is not much influenced by the RC> pulse duration or collection time by varying the interstrip capacitances from 1.2 to 1.5 pF/cm as per our previous published observations in the hadron/mixed irradiated detectors.

Based on the findings, a radiation hard ac coupled 200 μm n-Fz DSSSD design featuring an optimized intra guard ring on the back side of the detector and a WGR at the cut edge is proposed as an optimal design for the phase 1 upgrade of the R3B experiment.

## Acknowledgements

The author would like to give especial thanks to Prof. (Dr.) Kirti Ranjan and Prof. (Dr.) Ashutosh Bhardwaj, Centre for Detector and Related Software Technology, Department of Physics and Astrophysics, University of Delhi, New Delhi, India for providing necessary R & D infrastructure for the work.